\documentclass[10pt,journal,compsoc]{IEEEtran}
\usepackage{cite}
\usepackage{amsmath,amssymb,amsfonts}
\usepackage{algorithmic}
\usepackage{graphicx}
\usepackage{textcomp}
\usepackage{algorithm}
\usepackage{algorithmic}
\usepackage{hyperref}

\begin{document}
%
\title{ SDN-based Runtime Security Enforcement Approach for Privacy Preservation of Dynamic Web Service Composition }
%
\author{Yunfei~Meng,
        Zhiqiu~Huang,
        Guohua~Shen
        and~Changbo~Ke
}

\markboth{}%
{Yunfei Meng \MakeLowercase{\textit{et al.}}: SDN-based Runtime Security Enforcement Approach for Privacy Preservation of Dynamic Web Service Composition}

\IEEEtitleabstractindextext{%
\begin{abstract}
Aiming at the privacy preservation of dynamic Web service composition, this paper proposes a SDN-based runtime security enforcement approach for privacy preservation of dynamic Web service composition. The main idea of this approach is that the owner of service composition leverages the security policy model (SPM) to define the access control relationships that service composition must comply with in the application plane, then SPM model is transformed into the low-level security policy model (RSPM) containing the information of SDN data plane, and RSPM model is uploaded into the SDN controller. After uploading, the virtual machine access control algorithm integrated in the SDN controller monitors all of access requests towards service composition at runtime. Only the access requests that meet the definition of RSPM model can be forwarded to the target terminal. Any access requests that do not meet the definition of RSPM model will be automatically blocked by Openflow switches or deleted by SDN controller, Thus, this approach can effectively solve the problems of network-layer illegal accesses, identity theft attacks and service leakages when Web service composition is running. In order to verify the feasibility of this approach, this paper implements an experimental system by using POX controller and Mininet virtual network simulator, and evaluates the effectiveness and performance of this approach by using this system. The final experimental results show that the method is completely effective, and the method can always get the correct calculation results in an acceptable time when the scale of RSPM model is gradually increasing.
\end{abstract}

\begin{IEEEkeywords}
privacy preservation, SDN, Web service, access control, virtual machine.
\end{IEEEkeywords}}

\maketitle

\section{Introduction}
$\ $

\IEEEPARstart{W}{eb} service composition has been widely used in various fields of life. Users need to provide some personal information to service providers to support the implementation of functions provided by Web service composition. Some information involve the privacy of users. The privacy information here involves any information that is not publicly released by an individual user or enterprise, including personal identification information, health information, financial information, business secret or intellectual property rights. Because privacy information often contains potential huge commercial value, malicious service providers may illegally use or disclose these sensitive information without user permission under the interest driven, which will cause great harm to individual users or enterprises [1]. According to the survey released by Risk-based Security (RBS) in 2019, the global data disclosure events show an increasing trend year by year, of which the number of leakage events in 2019 is 33.3$\%$ higher than that in 2018, while that in 2019 is 112$\%$ higher than that in 2018. There are 5183 data disclosure events reported in the world from January 2019 to September 2019, most of which involve personal users or enterprise privacy information, and the amount of data leaked reached 7.7995 billion records.

At present, the main reasons for privacy leakage of Web service composition are the lack of effective privacy protection mechanism and illegal access of network layer. On the one hand, in the existing cloud computing environment, users can not evaluate the privacy security of Web services, and can not find web services that can meet their privacy needs according to their privacy preferences. This is because the current web service description mechanism based on Web service description language (WSDL) [4] can only describe the functional attributes of the service (such as service interface name, input and output variables of the interface, variable type, etc.), but cannot describe the non functional attributes of the service (such as security policy or privacy policy, etc.) [5]. On the other hand, the existing runtime monitoring mechanism of web service composition (such as BPEL execution engine, information flow monitor or SOAP message engine) can not automatically map the access control relationship that service composition must satisfy in the application layer to the network layer. As shown in Figure 1, the web service set {S1, S2, S3} constitutes a web service composition in the application layer, and SC1 is the service consumer of the web service composition; If the member service S2 cannot access the member service S3 in the application layer, the service consumer SC1 cannot access the member service S3; In the absence of security enhancement mechanism in the network layer, the network terminal T2 mapped by S2 and the Network Terminal T3 mapped by service S3 can still be connected in the network layer, and the network terminal T4 mapped by SC1 and the Network Terminal T3 mapped by service S3 can also be connected in the network layer; Therefore, the attacker can bypass the runtime monitoring mechanism of service composition deployed by users in the application layer, launch network-layer illegal access or identity theft attacks against the physical or virtual machine where the target service is located in the network layer, and steal its privacy information. Although network firewall can prevent the network-layer illegal accesses, but the existing network firewall mechanism does not have application-aware abilities. That is to say, when Web service composition dynamically evolves due to the change of application-layer requirements, the access control rules in the firewall cannot evolve synchronously and dynamically without human interventions, Therefore, the existing network firewall mechanism can not effectively solve the security enhancement problems of dynamic Web service composition.

\begin{figure}[htbp]
\centering
\scalebox{0.40}{\includegraphics{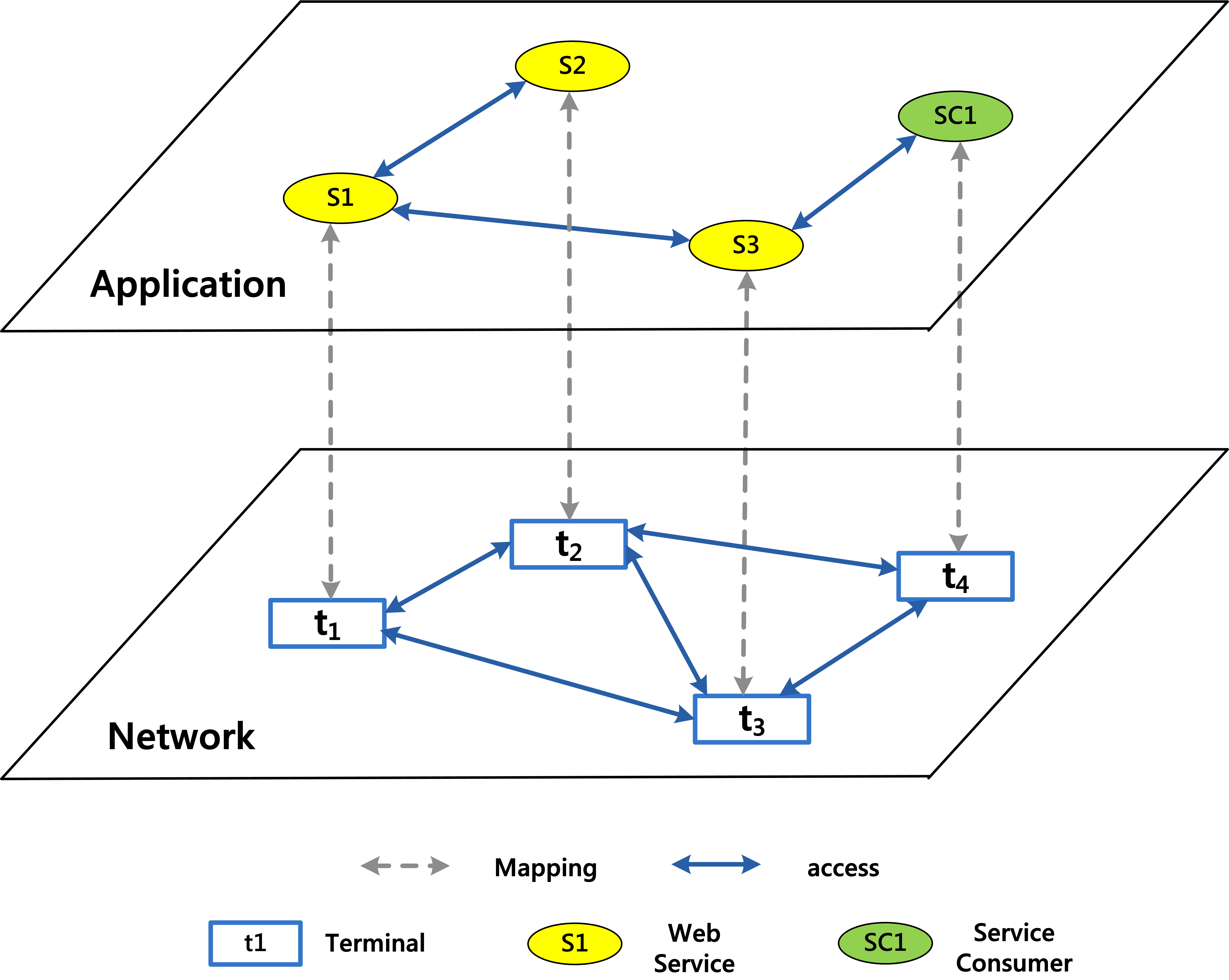}}
\caption{Traditional Network.}
\end{figure}

Towards the privacy preservation of Web service composition at runtime, we propose a SDN-based security enforcement approach for privacy preservation of dynamic Web service composition. The remainder of the paper is structured as follows. Section II is preliminary, Section III is problem description including of system model and attack model. Section IV is the main body of this paper, which presents our proposed framework in detail. Section V implements the framework and evaluates its effectiveness and performance. Section VI discusses some related works and compares them with our framework. Finally, Section VII concludes this paper and presents some future directions.

\section{Overview of SDN}
$\ $

Software-defined networking (SDN) is an approach which facilitates network management and enables more efficient network configurations \cite{SDN2}. The framework of SDN can be decomposed as application plane, control plane and data plane \cite{SDN1}. SDN suggests to centralize network intelligence in one network component by decoupling the controlling process (control plane) from the forwarding process (data plane). Openflow switch is the core of data plane, which consists of three parts: flow table, Openflow protocol and secure channel. The architecture of Openflow switch can be depicted in Figure 1.

\begin{figure}[htbp]
\centering
\scalebox{0.32}{\includegraphics{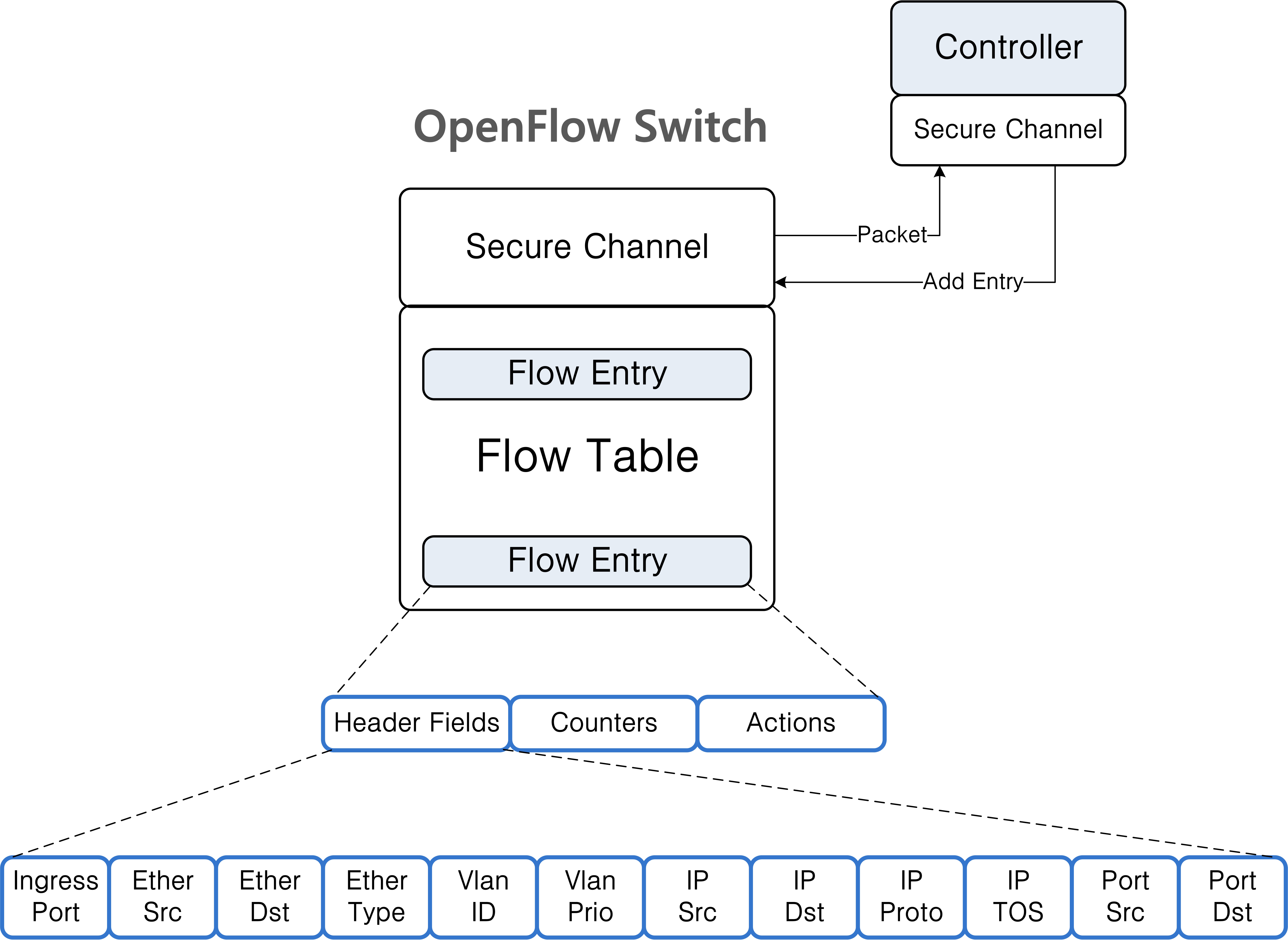}}
\caption{Architecture of Openflow Switch.}
\end{figure}

In this configuration, when a new packet arrives through in-ports, switch compares the header field of the packet against all flow entries in flow table. If this packet matches an existed entry, the switch updates its counters and executes the associated actions of entry, otherwise the packet will be sent to SDN controller through secure channel. By means of preloaded applications, the controller decides how to deal with the incoming packets. If packets can to be forwarded, controller inserts a new entry into flow table with relevant information of the packet, then forwards it. If packets need to be dropped, controller first clear counter of the packet, then drops it directly. Openflow Protocol regulates the format of information flows transmitted between Openflow switch and controller. The format of flow entry is defined as three parts: header fields, counters and actions. As shown in Figure 1, each header field can be further decomposed as 12 subfields, such as source IP, destination IP and some others.

Controller plane consists of three parts: secure channel, network operating system (NOS) and network applications. Secure channel is the interface in which controller can communicate with switch. Network applications are deployed on NOS platform, and implement some customized security or performance policies towards network management, such as firewall, VM migrations, intrusion detections and etc. Currently mainstream SDN controllers include open-source POX\cite{POX}, NOX\cite{NOX}, Floodlight\cite{Flood} and some other commercial controllers, such as NSX controller \cite{NSX} of VMware.

\section{Threat Models and Countermeasures}
$\ $

In this section, we first discuss the main threat models which the privacy preservation of Web service composition will confront at runtime, then we present the SDN-based countermeasures we adopt in this paper towards the threat models we have listed.

\subsection{Threat Models}
$\ $

In 2019, the cloud security alliance (CSA) published 12 top security threats [3], among which the security threats related to the privacy disclosure of Web services include the following three categories:

\textbf{Network-layer Illegal Access:} In traditional cloud computing architecture, the access control relationships defined in application plane cannot be mapped to the network plane automatically, so that the attacker can bypass the runtime security enforcement mechanisms deployed by users in application plane, such as information flow monitor or BPEL execution engine, directly access the physical or virtual machine of target Web service in network plane and steal its privacy data.

\textbf{Identity Theft Attack:} Identity theft attack refers to an attacker illegally access user's Web service composition and steal its privacy data in application plane by intercepting the personal identification information (identities or private key) of legitimate service consumers, employees or system managers \cite{Identity}.

\textbf{Service Leakage:} Service leakage refers to the malicious service provider illegally access user's Web service in application plane by using some special hidden service interfaces, which can bypass the security authentication mechanism set by user and steal its privacy data illegally \cite{insider}.

\subsection{SDN-based Countermeasures}
$\ $

Because the data plane of SDN has application-aware abilities, the countermeasures adopted in this paper is to enforce the security of Web service composition by means of SDN techniques, so as to solve the problems of illegal network accesses, identity theft attacks as well as service leakages existed in current cloud computing environment.

Specifically, a system model is proposed to describe the elements of SDN application plane and data plane in a formal way. Then, the relationship between Web services and network terminals is defined in the system model, and the relationship between service consumers and network terminals is also defined. Next, the access control relationships which service composition must comply with are defined as a security policy model (SPM) by user. Based on the model transformation method we proposed in [C and S], SPM model is transformed into the underlying security policy model (RSPM), which defines the access control relationships of network terminals which are associated with Web service composition in SDN data plane. In this way, we automatically map the access control relationships defined in SDN application plane into the access control relationships between network terminals in SDN data plane. Based on RSPM model, the access control program in SDN control plane authenticates each access request towards Web service composition at runtime. If finding the request doesn't comply with RSPM model, access control program deletes this request directly; If finding it conforms to the definitions of RSPM model, then convert the current RSPM model into a group of forwarding rules used by Openflow switches, then forwards this request after loading the generated forwarding rules into Openflow switches. Moreover, we have proofed that the forwarding rules generated from RSPM model can meet all of security properties defined in SPM model strictly in [C and S]. After loading the generated forwarding rules into SDN data plane, when attacker attempts to launch illegal access in SDN data plane or identity theft attacks in SDN application plane, when malicious service provider attempts to illegally access Web services using the hidden service interfaces, these access requests that do not comply with the definitions of RSPM model will be automatically blocked by Openflow switches, thus realizing to enforce the security of Web service composition at runtime.

\begin{figure}[htbp]
\centering
\scalebox{0.42}{\includegraphics{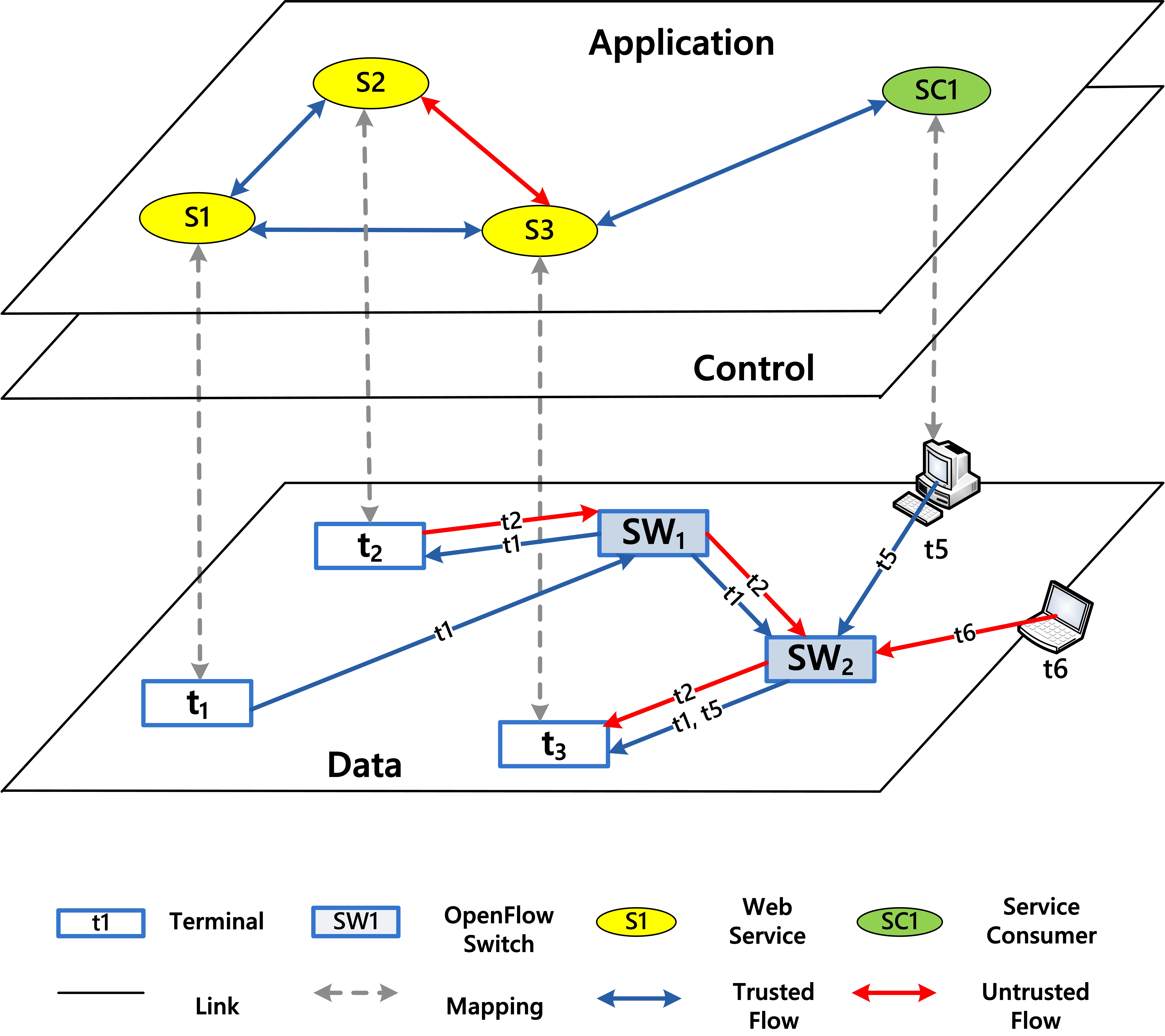}}
\caption{Countermeasures based on SDN.}
\end{figure}

As shown in Figure 2, the network terminals associated with Web Service $s_1$, $s_2$  and $s_3$ are $t_1$, $t_2$  and $t_3$ respectively, and the network terminal associated with service consumer $sc_1$ is $t_5$. In application plane, $\{$ $s_1$, $s_2$, $s_3$ $\}$ compose of a Web service composition (WSC). The access control relationship of SC is defined as follows: service $s_1$ can access $s_2$ and $s_3$ but $s_2$ cannot access $s_3$;  service consumer $sc_1$ can only access service $s_3$. Above access control relationships are transformed into a group of forwarding rules and loaded into the Openflow switch $sw_1$ and $sw_2$ respectively. After that, service $s_1$ can access $s_2$ and $s_3$ using terminal $t_1$; service consumer $sc_1$ can access service $s_3$ using terminal $t_5$. But when attacker launches illegal network accesses towards $s_3$ using terminal $t_2$ in data plane, such access request will be blocked because it doesn't comply with the forwarding rules of switch $sw_1$; when attacker launches identity theft attacks towards $s_3$ using terminal $t_6$, such access request also will be blocked because it doesn't comply with the forwarding rules of switch $sw_2$. Thus, the security of Web service composition can be effectively enforced.

\section{System Model}
$\ $

In order to transform the access control relationships defined in SDN application plane into the forwarding rules used in Openflow switches, we propose a group system models which can formally describe the objects we research on and the mapping relationships between the application plane objects and data plane objects. In the following, we first establish the precise formal definitions of system model, then utilize a practical example to illustrate how to construct the system model of Web service composition.

\subsection{Formal Definition}
$\ $

\textbf{Terminal:} The network terminal existed in SDN data plane can be any physical or virtual machine which has computing and networking capabilities. The terminals are formally defined as a set $TM$=$\{$ $t_0$, $t_1$,..., $t_n$ $|\ \forall t_{i}$=$\langle$ $ip_i$, $mac_i$ $\rangle$ $\}$, where:$\\$
$\bullet\ ip_{i}$ represents the terminal's IP address used in data plane. We define each terminal can only have one unique IP address, i.e., if $t_{i}\neq t_{j}$, $ip_i\in t_i$, $ip_j\in t_j$, then $ip_{i}\neq ip_{j}$ must be held.$\\$
$\bullet\ mac_{i}$ represents the terminal's MAC address used in data plane. We define each terminal can only have one unique MAC address, i.e., if $t_{i}\neq t_{j}$, $mac_i\in t_i$, $mac_j\in t_j$, then $mac_{i}\neq mac_{j}$ must be held.

\textbf{Openflow Switch:} The Openflow switch existed in SDN data plane can be any physical or virtual switch based on Openflow protocol. Openflow switches are formally defined as a set $SWM$=$\{$ $sw_{0}$, $sw_{1}$,...,$sw_{n}$ $|\ \forall$ $sw_{i}$=$\langle$ $SP$, $FT$ $\rangle$ $\}$, where: $\\$
$\bullet$ $SP$=$\{$ $sp_{0}(sw_{i})$, $sp_{1}(sw_{i})$,...,$sp_{n}(sw_{i})$ $\}$ represents the set of ports exited in the switch $sw_{i}$.$\\$
$\bullet$ $FT$=$\{$ $tr_{0}$, $tr_{1}$,...,$tr_{n}$ $|\ \forall$ $tr_{i}$=$\langle$ $Src_{mac}$, $Src_{ip}$, $Dst_{mac}$, $Dst_{ip}$, $Inp$, $Outp$ $\rangle$ $\}$ represents the flow table of switch $sw_i$, which is a set of forwarding rules $\{ tr_i \}$, where $Src_{mac}$ represents the MAC address of the terminal sending package, $Src_{ip}$ represents the IP address of the terminal sending package; $Dst_{mac}$ represents the MAC address of the terminal receiving package; $Dst_{ip}$ represents the IP address of the terminal receiving package; $Inp\in SP$ represents the port receiving package; $Outp\in SP$ represents the port forwarding package. By default, we define $FT$ = $\varnothing$, which indicates the Openflow switch $sw_{i}$ cannot forward any packages.

\textbf{Network Topology:} The network topology of SDN data plane is formally defined as an undirected graph $NTM$=$\langle$ $V$, $C$, $E$ $\rangle$, where: $\\$
$\bullet$ $V\subseteq TM\cup SWM$ represents a finite set of network vertexes. $\\$
$\bullet$ $C$=$\{$ $c_{0}$, $c_{1}$,...,$c_{n}$ $|\ \forall$ $c_{i}\in\mathbb{R}^{+}$ $\}$ represents a finite set of the cost of network links.$\\$
$\bullet$ $E\subseteq P\times C\times P$ represents a finite set of network links; $P\subseteq TM\cup SP^{*}$, where $SP^{*}$ represents the universal set of switch ports existed in $SWM$. If $\exists t_{m}\in TM$, $\exists t_{n}\in TM$, $\exists sp_{m}(sw_{i})\in SP^{*}$, $\exists sp_{n}(sw_{i})\in SP^{*}$, $\exists sp_{m}(sw_{j})\in SP^{*}$, $\exists c_{i}\in C$, then $e_i$ = $\{$ $t_{m}$, $c_{i}$, $sp_{m}(sw_{i})$ $\}$ represents a network link from the terminal $t_{m}$ to the port $sp_{m}(sw_{i})$, whose cost equals $c_{i}$; $e_i$ = $\{$ $t_{m}$, $c_{i}$, $t_{n}$ $\}$ represents a network link from the terminal $t_{m}$ to the terminal $t_{n}$, whose cost equals $c_{i}$; $e_i$ = $\{$ $sp_{m}(sw_{i})$, $c_{i}$, $sp_{m}(sw_{j})$ $\}$ represents a network link from the port $sp_{m}(sw_{i})$ to the port $sp_{m}(sw_{j})$, whose cost equals $c_{i}$; $e_i$ = $\{$ $sp_{m}(sw_{i})$, $c_{i}$, $sp_{n}(sw_{i})$ $\}$ represents an inner network link from the port $sp_{m}(sw_{i})$ to the port $sp_{m}(sw_{i})$ in switch $sw_{i}$, in this case, we define the cost of any inner network link in Openflow switch equals Zero (i.e., $c_{i}=0$).

\textbf{Web Service:} Web services existed in SDN application plane are formally defined as a set $SM$=$\{$ $s_0$, $s_1$,..., $s_n$ $|\ \forall s_{i}$=$\langle$ $SPPM$, $t_i$, $su_i$ $\rangle$ $\}$, where:$\\$
$\bullet\ SPPM$ represents the privacy policy of Web service. $\\$
$\bullet\ t_{i}\in TM$ represents the terminal in SDN data plane, which is mapped with Web service. $\\$
$\bullet\ su_{i}$ represents the uniform resource identifier (URI) of Web service; if $s_{i}\neq s_{j}$, $su_i\in s_i$, $su_j\in s_j$, then $su_{i}\neq su_{j}$ must be held.

\textbf{Service Consumer:} The services consumer existed in SDN application plane is the user of Web service composition. Service consumers are formally defined as a set $CM$=$\{$ $c_0$, $c_1$,..., $c_n$ $|\ \forall c_{i}$=$\langle$ $CPPM$, $t_i$ $\rangle$ $\}$, where:$\\$
$\bullet\ CPPM$ represents the privacy preferences of service consumer. $\\$
$\bullet\ t_{i}\in TM$ represents the terminal in SDN data plane, which is mapped with service consumer.

\textbf{Web Service Composition:} Web service composition is formally defined as a finite state machine $WSC$ = $\langle$ $I_{0}$, $S$, $C$, $E$ $\rangle$, where:$\\$
$\bullet\ I_{0}\in SM$ represents the initial state of Web service composition.$\\$
$\bullet\ S\subseteq SM$ represents a finite set of Web services which compose Web service composition.$\\$
$\bullet\ C$ represents a finite set of invoking events.$\\$
$\bullet\ E\subseteq S\times C\times S$ represents a finite set of directed transitions. If $\exists s_{i}\in SM$, $\exists s_{j}\in SM$, $\exists c_{k}\in C$, then $e_i$ = $\{$ $s_{i}$, $c_{k}$, $s_{j}$ $\}$ represents a directed transition from service $s_i$ to service $s_j$, which means service $s_i$ invokes service $s_j$ under the invoking event $c_k$ having been triggered.

\subsection{Modelling Example}
$\ $

Smart physiological monitoring system (SPMS) is an application based on Web service composition and Internet of things (IoT) techniques. As shown in Figure 3, SPMS mainly consists of five critical parts: login service, monitoring service, alarming service, heart rate sensing service and temperature sensing service. The working process of SPMS can be described as follows. Only patient and attending physician defined by patient can pass the security authentication of login service, and read the real-time patient's physiological data using monitoring service. If finding patient's body temperature has exceeded the threshold, temperature sensing service connected with patient's temperature sensors will send an alert message (te=1) to the monitoring service, otherwise it sends the message (te=0). If finding patient's heart rate has exceeded the threshold, heart rate sensing service connected with heart rate sensors will send an alert message (hr=1) to the monitoring service, otherwise it sends the message (hr=0). If finding $hr\cup te$=1, monitoring service will invoke the alarming service to remind attending physician of patient.

\begin{figure}[htbp]
\centering
\scalebox{0.55}{\includegraphics{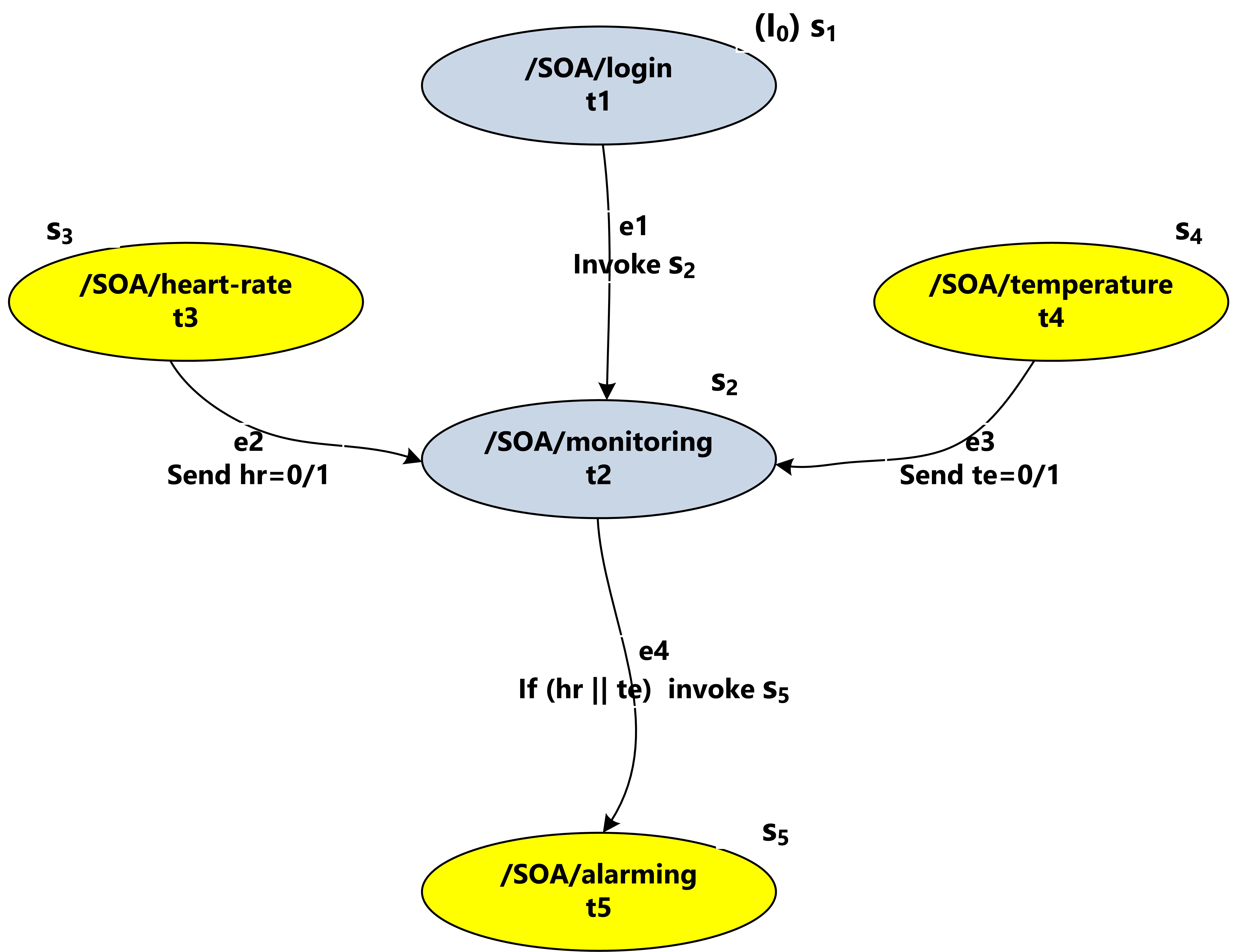}}
\caption{Web service composition of SPMS. Gray components represent application services, yellow components represent sensing services.}
\end{figure}

According to above descriptions, login service of SPMS is formally defined as $s_{1}$=$\langle$ $SPPM$, $t_{1}$, $/SOA/login$ $\rangle$; monitoring service of SPMS is formally defined as $s_{2}$=$\langle$ $SPPM$, $t_{2}$, $/SOA/monitoring$ $\rangle$;  heart rate sensing service is formally defined as $s_{3}$=$\langle$ $SPPM$, $t_{3}$, $/SOA/heart\ rate$ $\rangle$; temperature sensing service of SPMS is formally defined as $s_{4}$=$\langle$ $SPPM$, $t_{4}$, $/SOA/temperature$ $\rangle$; alarming service of SPMS is defined as $s_{5}$=$\langle$ $SPPM$, $t_{5}$, $/SOA/alarming$ $\rangle$ represents the. Web service composition of SPMS is formally defined as $WSC$ = $\langle$ $I_{0}$, $S$, $C$, $E$ $\rangle$, where:$\\$
$\bullet\ I_{0}$=$s_1$.$\\$
$\bullet\ S$=$\{$ $s_1$, $s_2$, $s_3$, $s_4$, $s_5$ $\}$. $\\$
$\bullet\ C$=$\{$ $c_1$, $c_2$, $c_3$, $c_4$ $\}$. $\\$
$\bullet\ E$=$\{$ $e_1$=$\{$ $s_1$, $c_1$, $s_2$$\}$, $e_2$=$\{$ $s_3$, $c_2$, $s_2$$\}$, $e_3$=$\{$ $s_4$, $c_3$, $s_2$$\}$, $e_4$=$\{$ $s_2$, $c_4$, $s_5$$\}$ $\}$.

\section{The Approach}
$\ $

Based on the established system model, we propose a SDN-based runtime security enforcement approach for privacy preservation of dynamic Web service composition in this paper. In the following of this section, we first overview the framework of this approach, then present how to specify the access control relationships complied with Web service composition in application plane using security policy model (SPM), as well as how to transform SPM model into the low-level security policy model (RSPM). Finally, we present the virtual machine access control algorithm integrated in the SDN controller.

\subsection{Overview of Approach }
$\ $

The framework of this approach is depicted as Figure 4. Specifically, this approach is decomposed as application plane, control plane and data plane. First of all, the access control relationships defined by the owner of Web service composition are specified as a formal security policy model (SPM) by the security policy model construction algorithm. Based on the system model proposed in this paper, the low-level security policy model construction algorithm transforms SPM model into a formal low-level security policy model (RSPM) which defines the access control relationships between network terminals in data plane. In this way, the access control relationships of Web service composition defined in application plane have been transformed into the corresponding access control relationships between network terminals in data plane. After uploading RSPM model into control plane (i.e., SDN controller), any access requests complying with the definition of RSPM model will be forwarded by Openflow switches directly; any access requests sent from the malicious service consumers defined by the owner will be blocked by Openflow switches directly; when an access request sent from an unknown service consumer arrives at the Openflow switch in data plane, the switch will send a query request (i.e., packet-in packet) to the SDN controller for judgment. The virtual machine (VM) access control algorithm integrated in the SDN controller first extracts the source terminal information $ST\in TM$ and target terminal information $DT\in TM$ from the header fields of packet-in packet, then reads the latest uploaded RSPM model and compares the extracted the pair $\langle$ $ST$, $DT$ $\rangle$ with the information of RSPM model. If the pair meets the definition of RSPM model, VM access control algorithm converts the current RSPM model into a group of forwarding rules for Openflow switch, then permits the access request of service consumer by updating all of Openflow switches with the generated forwarding rules; If the pair doesn't meet the definition of RSPM model, VM access control algorithm denies the access request of service consumer by deleting this packet-in packet directly. In this way, when attacker attempts to launch illegal network accesses in data plane or identity theft attacks in application plane, when malicious service provider attempts to illegally access Web services using the hidden service interfaces, these access requests that don't comply with the definition of RSPM model will be blocked by Openflow switches or be deleted by the SDN controller, thus this approach can effectively enforce the security of Web service composition at runtime.

\begin{figure}[htbp]
\centering
\scalebox{0.45}{\includegraphics{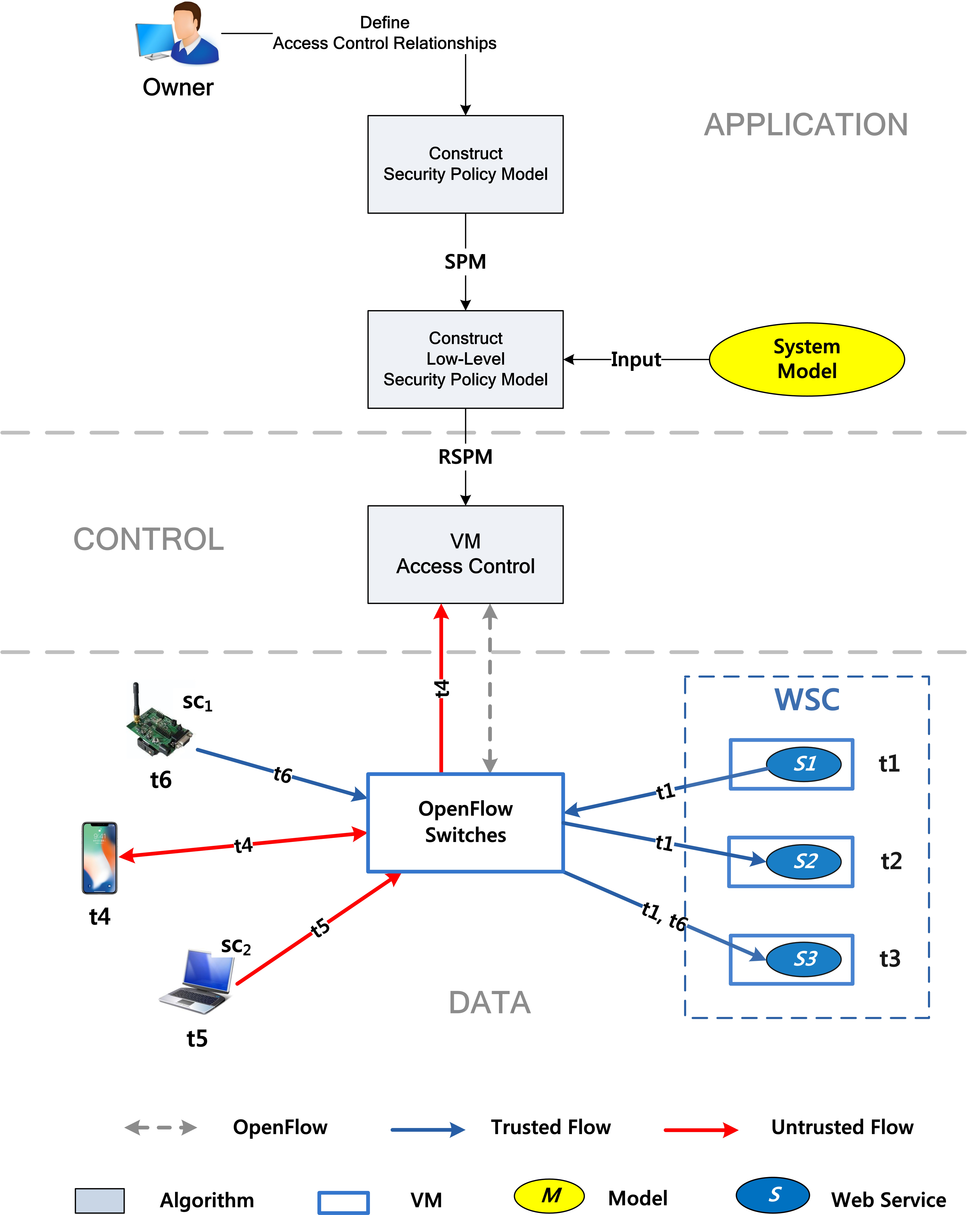}}
\caption{The framework of approach. }
\end{figure}

As shown in Figure 4, Web services $\{$ $s_1$, $s_2$, $s_3$ $\}$ compose a Web service composition (WSC) in application plane. In data plane, service $s_1$ runs in VM $t_1$, service $s_2$ runs in VM $t_2$ and service $s_3$ runs in VM $t_3$. The owner of WSC defines the access control relationships using SPM model, i.e., service $s_1$ can access service $s_2$ and $s_3$; service consumer $sc_1$ can access service $s_3$; service consumer $sc_2$ is a malicious consumer. After transforming SPM model into RSPM model and uploading RSPM model into the SDN controller, in data plane, VM $t_1$ of service $s_1$ can only access VM $t_2$ of service $s_2$ and VM $t_3$ of service $s_3$; sensor $t_6$ of service consumer $sc_1$ can only access VM $t_3$ of service $s_3$; any access requests sent from the terminal $t_5$ which is associated with the malicious consumer $sc_2$ will be blocked by Openflow switch directly; the access request sent from the terminal $t_4$ which is associated with an unknown service consumer will be forwarded into the SDN controller by Openflow switch, then it will be deleted by VM access control algorithm in the SDN controller because it doesn't comply with the definition of RSPM model.

\section{Related Work}
$\ $

The SDN-based gateway of our framework provides a firewall mechanism which guarantees only authorized things can access service providers's virtual machine, while those undefined things can not. Hence, in this section, we want to discuss some research works concerning how to implement dynamic firewall mechanism using SDN, and compare these proposals with our framework.

Hu et al. \cite{SDN9} proposed a comprehensive framework, Flowguard, to facilitate accurate detection as well as flexible resolution of firewall policy violations in dynamic Openflow networks. In addition, authors implemented a prototype using Floodlight. The experimental results show that Flowguard has the manageable performance overhead to enable realtime monitoring network. Similarly, Porras et al. \cite{SDN5} proposed a security enforcement controller, FortNOX, which is an extension on NOX controller. FortNOX is designed to enable a network flow to be blocked (or allowed) by security applications. They also proposed a conflict resolving mechanism used in case of appearing policy conflicts. Exactly, we are inspired by the ideas of Flowguard and FortNOX in some sense, we also design the relevant policy resolving mechanism in our framework, i.e., the information flow rules of administrator ($Role=A$) can override those rules of users ($Role=U$). Moreover, we design all entries in OVS can be automatically updated per $t$ minutes, which can also be used to resolve the policy conflicts.

Suh et al. \cite{SDN7} leveraged POX controller to implement a firewall application. Each firewall rule can be defined by 6 actions and 12 conditions, and the final experimental results illustrate the firewall is effective. But this mechanism requires network operators to know the details of underlying network, and input the firewall rules into the controller manually. While in our framework, all of information flow rules of IFM are converted from SRM automatically, service providers just need to know which service could be released to which consumer or which thing, other details of underlying network can be created from system models automatically. Therefore, any normal user can leverage our framework to rapidly define their security policies.

Koerner et al. \cite{SDN6} proposed a MAC-based VLAN tagging mechanism using SDN. The virtual local area network (VLAN) has been widely used in enterprise networks where the security policy is always defined by VLAN address. But some mobile laptop-based workstations often change their locations, which will leads to the frequent changing of its VLAN address and incur security policy conflicts. To address this problem, authors leverage Floodlight controller to map the MAC address of laptop into its corresponding VLAN address in network. Since MAC address is static, thus it can guarantees the laptops can access the network successfully in different locations. In our framework, the controller use information flow rule to recognize an authorized user, i.e., the pair $\langle$ $Src$, $Dst$ $\rangle$. Here $Src$ is MAC address of service consumer, $Dst$ is VLAN address of VM, but we don't need to convert MAC address into a VLAN address.

In addition, Javid et al. \cite{SDN8} implemented a 2-layer firewall using POX controller. CloudWatcher \cite{SDN3} is a security monitoring framework by which network operators can define a policy to describe a network traffic and describe which security services must be applied to it. Koorevaar et al. \cite{SDN4} proposed an framework for leveraging SDN for automatic security policy enforcement using EEL-tags. These tags are added into the VM's flow by hypervisor. By means of these added EEL tags, they can implement the associated security policy. However, this work heavily relies on trustful hypervisor, thus the portability of method is a big problem need to be considered.

\section{Conclusion}
$\ $

Aiming at the privacy preservation of dynamic Web service composition, this paper proposes a SDN-based runtime security enforcement approach for privacy preservation of dynamic Web service composition. The main idea of this approach is that the owner of service composition leverages the security policy model (SPM) to define the access control relationships that service composition must comply with in the application plane, then SPM model is transformed into the low-level security policy model (RSPM) containing the information of SDN data plane, and RSPM model is uploaded into the SDN controller. After uploading, the virtual machine access control algorithm integrated in the SDN controller monitors all of access requests towards service composition at runtime. Only the access requests that meet the definition of RSPM model can be forwarded to the target terminal. Any access requests that do not meet the definition of RSPM model will be automatically blocked by Openflow switches or deleted by SDN controller, Thus, this approach can effectively solve the problems of network-layer illegal accesses, identity theft attacks and service leakages when Web service composition is running. In order to verify the feasibility of this approach, this paper implements an experimental system by using POX controller and Mininet virtual network simulator, and evaluates the effectiveness and performance of this approach by using this system. The final experimental results show that the method is completely effective, and the method can always get the correct calculation results in an acceptable time when the scale of RSPM model is gradually increasing.

\section*{Acknowledgment}
$\ $

This paper has been sponsored and supported by National Natural Science Foundation of China (Grant No.61772270), partially supported by National Natural Science Foundation of China (Grant No.61602262).


\bibliographystyle{IEEEtran}
\bibliography{test}


%

\end{document}